\documentclass[5p,times,authoryear]{elsarticle}

\usepackage{ecrc_RIAI}
    
\usepackage[utf8]{inputenc}
\usepackage[english]{babel}

\addto\captionsspanish{%
}

\usepackage{amsmath}            
\usepackage{epstopdf}           
\usepackage{flushend}           

\volume{00}

\firstpage{1}

\journalname{Cities}

\runauth{Primer autor et al.}

\jid{RIAI}

\jnltitlelogo{}

\usepackage{amssymb}

\usepackage[figuresright]{rotating}

\usepackage{subfigure}

\usepackage{url}

\usepackage{breakurl}
\usepackage[breaklinks]{hyperref}

\begin{document}

\begin{frontmatter}






\title{The ubiquitous efficiency of going further: how street networks affect travel speed}


\author[First]{Gabriel L. Maia}

\author[First]{Caio Ponte}

\author[First]{Carlos Caminha}

\author[First]{Lara Furtado}

\author[Second]{Hygor P. M. Melo}

\author[First]{Vasco Furtado}

\address[First]{University of Fortaleza Av. Washington Soares, 1321 - Bl J Sl 30, 60.811-905, Fortaleza, Brazil.}
\address[Second]{Centro de F\'{i}sica Te\'{o}rica e Computacional, Faculdade de Ci\^{e}ncias, Universidade de Lisboa, 1749\textendash016 Lisboa,
Portugal.}
\address[Third]{University of Masschusetts Amherst, Amherst, MA 01003, USA}

\begin{abstract}

As cities struggle to adapt to more “people-centered” urbanism, transportation planning and engineering must innovate to expand the street network strategically in order to ensure efficiency but also to deter sprawl. Here, we conducted a study of over 200 cities around the world to understand the impact that the patterns of deceleration points in streets due to traffic signs has in trajectories done from motorized vehicles. We demonstrate that there is a ubiquitous nonlinear relationship between time and distance in the optimal trajectories within each city.  More precisely, given a specific period of time $\tau$, without any traffic, one can move on average up to the distance $\left \langle D \right \rangle \sim\tau^\beta$. We found a super-linear relationship for almost all cities in which $\beta>1.0$. This points to an efficiency of scale when traveling large distances, meaning the average speed will be higher for longer trips when compared to shorter trips. We demonstrate that this efficiency is a consequence of the spatial distribution of large segments of streets without deceleration points, favoring access to routes in which a vehicle can cross large distances without stops. These findings show that cities must consider how their street morphology can affect travel speed. 

\end{abstract}

\end{frontmatter}


\section{Introduction}

The growth of human mobility enabled by recent urbanization has created prosperous economical and social urban centers \cite{bettencourt2007growth,grimm2008global}. On the other hand, urbanization can also have detrimental effects and poses particular challenges to studies in mobility, traffic and sustainability \cite{boltze2016approaches,rutledge2010thought}. One such issue is that of urban sprawl, where urbanization takes place at a fast pace consuming rural and green areas with no specific planning, mostly guided by speculative practices \cite{Laidley2016}. The expansion of the urban fabric brought by sprawl requires an understanding of how to build efficient street networks to absorb new vehicles coming from low density areas.

Contemporary planning, often led by New Urbanism principles, has staked claim against sprawl, in favor of downtown areas, walkable neighborhoods, smaller building blocks and denser concentric cities in order to produce more sustainable and traffic-free environments \cite{Ewing2018,calthorpe2000new}. An understanding of individual and collective human mobility is a particularly important aspect to understand the historical evolution of cities and social aspects of urban life \cite{makse1995modelling,bettencourt2007growth,louf2014congestion,caminha2017human}. It is essential for urban planning to design urban networks that can improve wellbeing \cite{ersoy2016landscape,kraemer2020effect,caminha2018graph,furtado2017data} and optimize transport structures \cite{liu2017intelligent,huang2018modeling, caminha2016micro, ponte2018traveling,mastroianni2015local,biazzo2019general}. However, in the sprawl debate, it remains unclear to which extent the expansion of cities and its street network are necessarily linked to poor traffic \cite{Ewing2018}. Commuting time and speed also can be related multiple factors such as the location of employment hubs \cite{Crane2003}, housing centrality or density \cite{Sarzynski2006}.

The complexity of the phenomenon of mobility has prompted studies that investigate how multi-modal transport combine individual vehicles, bicycles, buses and even airplanes to optimize human mobility \cite{gallotti2014anatomy}. Certain modes of transport can make longer trips more time-efficient: moving from New York to London by plane takes less hours per $km$, than moving from New York to Boston by car, for instance. In this example, the typical time to complete the trip does not increase proportionally with the distance, defining a non-linear time gain, which is a natural consequence of different modal speeds \cite{varga2016further}. Recent studies reveal that the same non-linearity can also be found on the time-distance scaling for urban travel even when using the same mode of transport \cite{varga2016further, gallotti2016stochastic,mastroianni2015local,biazzo2019general}. Even though dwellers living in suburban areas commute larger distances, their travel time is in fact shorter (higher average speed) than that of those coming from an urban downtown \cite{Kahn2006}. When driving is concentrated in more compact areas it also leads to an increase in travel time, which points to the importance of increasing street surface \cite{Ewingetal2018}.

The relationship between the street network and commuting time is particularly relevant for this study. Jayasinghe et al. \cite{Jayasinghe2019} developed a methodology to model vehicular traffic volume by using information on road segments based on the notion of network centrality. Their method did not require extensive and expensive datasets commonly used in traffic modeling such as OD and vehicle tracking data. However, the authors point to the study limitations since the method was not sufficient to model traffic volume at a macro level. 

Another study of street segments and traffic load over time used GPS car data to identify geographic clusters based on driver's behavior \cite{Necula2015}. After crossing those clusters with speed values, the authors found that greater road density implies lower average speed. \cite{Loder2019} investigated how network topology determines the network system's critical point, defined as the maximum number of vehicles in circulation while maintaining a continuous flow. Since traffic planning seeks to operate at a critical point, the author apply concepts of centrality to maximize travel speed. The study was conducted with 40 cities worldwide, and they find a sub-linear relationship between network size and critical accumulation. This means that the smaller the network size, the larger the accumulation of vehicles in an urban network \cite{Loder2019}. 

Galloti and Barthelemy suggests that speed fluctuations in a car's trajectory is what explains the non-linear relation between time and distance of trips \cite{gallotti2016stochastic}. This brings us to the important concept of free flow traffic conditions defined as "conditions where vehicular traffic is typically characterized by low to medium vehicular density, arbitrarily high mean speeds, and stable flow, over uninterrupted roadway segments" \cite{Khabbaz2012}. Ensuring free flow is a challenge in and of itself determined by density, flow and speed \cite{Khabbaz2012}. These variables have been assessed in different cases with varying objectives.

In Sweden, for instance, the Government introduced a traffic policy that reduced permitted speed limits in order to lower the overall traffic speed and the severity of traffic accidents. Simulations of the policy effects found that road characteristics and road environment such as design changes, sidewalks, and speed bumps were more efficient than actual posted speed limitations in impacting free flow speed \cite{Silvano2020}. Physical conditions of the street network such as street connectivity can promote free flow traffic and reduce travel times and delays \cite{Zlatkovic2019}.

There is still a lack of studies on how the expansion of the urban street footprint, street signage and capacity can influence trip efficiency of motorized vehicles within cities. This research contributes to that gap and presents an empirical experiment that simulates vehicle trips in order to measure time and distance of trips considering the existing street layout. Specifically, we employ data from the street network to investigate how the spatial distribution of deceleration points - \emph{DP} (i.e. points where vehicles can potentially stop or lower speed) and the street typology (correlated with specific permitted speeds) can influence trips' time and distance.

We conducted simulations of motorized trips for 228 cities worldwide by taking into account only variables pertaining to the streets' built properties such as the existence of traffic lights and stops, street intersections and permitted driving speed for each type of road (such as highway, motorway, residential, to name a few). We did not consider dynamic aspects such as traffic and vehicle flow from specific origin-destination travel patterns and demands. We found that, for a given time, the average distance traveled in each simulated trip increases proportionally faster than its duration. As an example, this means that a trip with double the time will typically travel for more than double of the distance.

We find that this phenomenon depends mostly on the amount of Segment Without Deceleration Points (\emph{SWDP}). We coined that term to refer to a segment of one or more streets that contain no points that force stopping or deceleration. The size, distribution and number of SWDP impacts a super-linear coefficient (described by a $\beta$ exponent), meaning that longer SWDP with good spatial coverage are easier to access and increase the gain of scale between the relation of time and distance. A particular empirical experiment we conducted supported that finding: when we simulated the same amount of trips and routes with randomly distributed SWDP we found the super-linearity to be completely absent.

It was also possible to classify cities based on their super-linear magnitude, average street speed and city size (built footprint or square footage). Mega-cities with extensive urban environments of thousands of squared km tend to present higher $\beta$ coefficients (around 1.20), while small cities present lower values (around 1.07). We find that some cities can have a good coverage of SWDP, but their smaller extension in terms of area do not allow for long SWDP, which leads to lower $\beta$  values (La Plata, Argentina). Other small cities with less SWDP, made up of smaller street segments, do not permit as much free flow and make that gain of scale less prominent (Copenhagen, Denmark). Larger cities known for having robust permeating highway systems also enable higher $\beta$ with either low average speeds (Paris, France) or high average speeds (Dubai, Arabic Emirates). In short, the distribution of SWDP and the urban footprint are the explaining factors for how cities fare in terms of its level of economy of scale when driving further.

\section{Methodology}

This empirical research analyzed cities and simulated optimal travel routes based on the street footprint and permitted speed and flows. The routes are completed by automotive vehicles and drawn within the existing street layout extracted from the open and collaborative mapping tool \textit{OpenStreetMap}. We did not include traffic data in the simulations and focused on the characteristics of the built street network such as signage, speed and intersections. To represent cities, we employed a directed graph $G (V,E)$, where nodes, $v$ $(\in V)$, indicate street intersections and edges, $e$ $(\in E)$, represent directed street segments between them. We used the library \textit{osmnx} to generate the graph based on \textit{OpenStreetMap} data \cite{boeing2017osmnx}. We extracted information on street lights, street typology (motorway, primary, secondary, etc.), maximum street speed and length of street segments for 228 cities in different continents.

To simulate the routes, we start with random origin points and generate routes in every possible direction, following the street typology, permitted direction flows and the deceleration points on the way. We run the trip until a certain time threshold is achieved and estimate the speeds based on the maximum speed provided by \textit{OpenStreetMap}. We then calculate the correlation between trip time and distance traveled for each optimal route in different origin points.

\subsection{Deceleration Points}

\begin{figure}[h!]
\centering
    \includegraphics[width=0.4\textwidth]{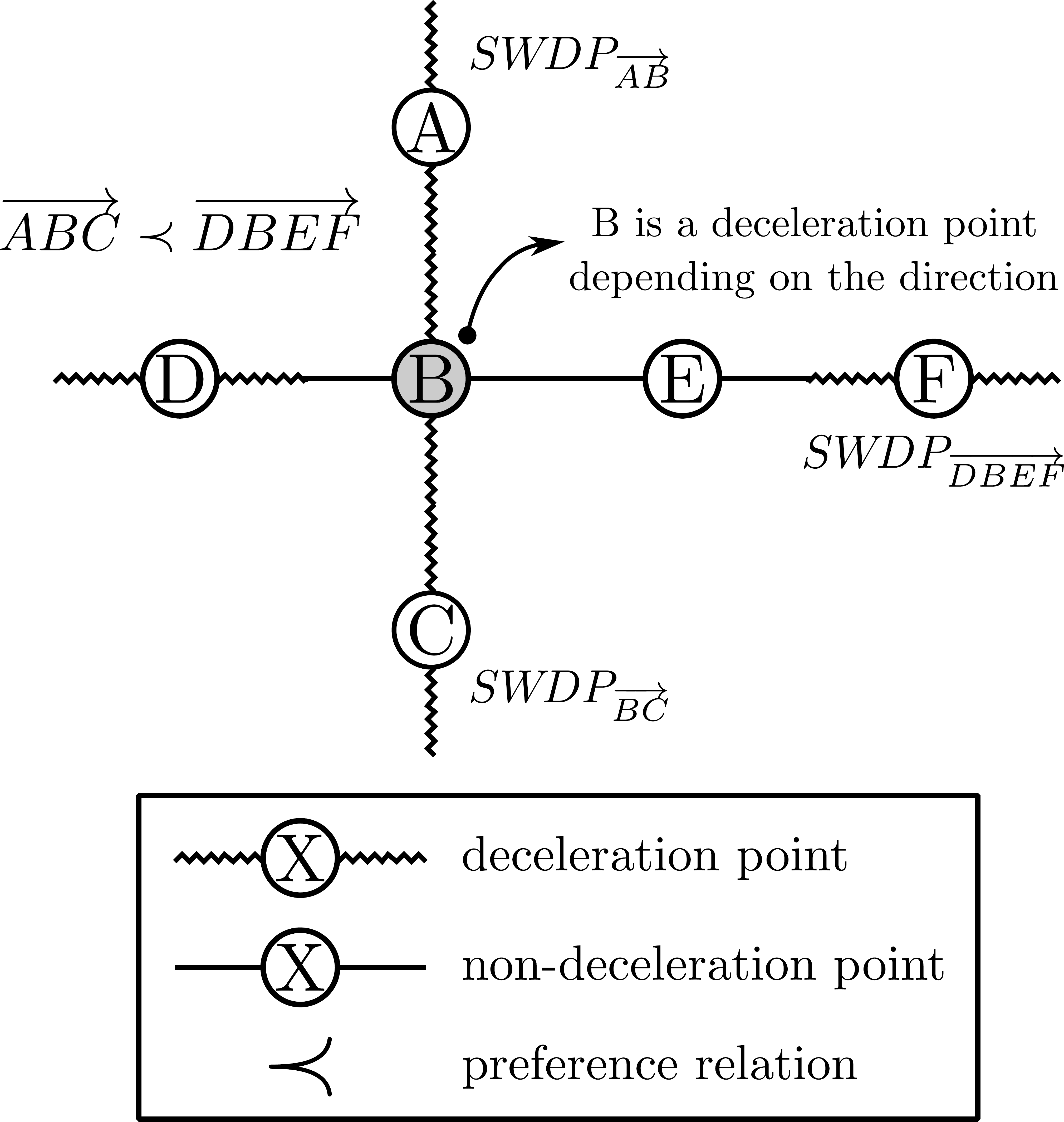}
     \caption{{\bf Identification of Deceleration Points (\emph{DP}) and Segments Without Deceleration Points (\emph{SWDP}).} The figure presents an example of how to identify a \emph{DP}. In the cross-shaped diagram, the circles represent crossroads - street intersections, while the lines represent segments connecting these nodes. The relation of preference given to each street, shown by ($\prec$), indicates which nodes to classify as \emph{DP}. This classification is dynamic and depends on the route the vehicle takes. For instance, the node B highlighted is classified as a \emph{DP} if the vehicle takes a route moving in direction $\protect\overrightarrow{ABC}$ while for the direction $\protect\overrightarrow{DBEF}$ the node B is not a \emph{DP}. This takes place because $\protect\overrightarrow{ABC} \prec \protect\overrightarrow{DBEF}$. The drawing style of lines in Figure help illustrate when node is a \emph{DP} (dashed lines) and not a \emph{DP} (filled lines). Once the \emph{DP} are established, the \emph{SWDP} can be defined by joining the segments between points. The example presents three different \emph{SWDP}: $SWDP_{\protect\overrightarrow{AB}}$, $SWDP_{\protect\overrightarrow{BC}}$ and $SWDP_{\protect\overrightarrow{DBEF}}$. The first and last nodes of the \emph{SWDP} will always be a \emph{DP} and the intermediate nodes, if existent, will not be \emph{DP}.}

\label{fig:mapa}
\end{figure}

Deceleration points ($DP$) are street nodes that can potentially cause stops or significant reductions in vehicle speed. They are not defined on the maps and are also not static since they change based on the direction of the route. Figure \ref{fig:mapa} shows part of a street network where the circles represent crossings and the lines represent streets. If the route begins from points $A$ or $C$, the node $B$ is highlighted as a \emph{DP}. If the vehicle begins a trip from point $D$, then it does not characterize as a $DP$. If the vehicle is moving from $D$ to $F$, it does not need to stop at $B$ because the segment $\overrightarrow{AB}$ has lower preference than segment $\overrightarrow{DBEF}$. It is also possible to define Segments Without Deceleration Points (\emph{SWDP}), which are when the intermediate segments of one or more streets are not \emph{DP}. Figure \ref{fig:mapa} shows three \emph{SWDP}: the $SWDP_{\overrightarrow{DBEF}}$, which starts at \emph {D} and ends at \emph {F}, the nodes $B$ and $E$ which are not \emph{DP}, and the segments $SWDP_{\overrightarrow{AB}}$ and $SWDP_{\overrightarrow{BC}}$ which present only a start and finish node.

\begin{figure*}[h!]
\centering
    \includegraphics[width=0.8\textwidth]{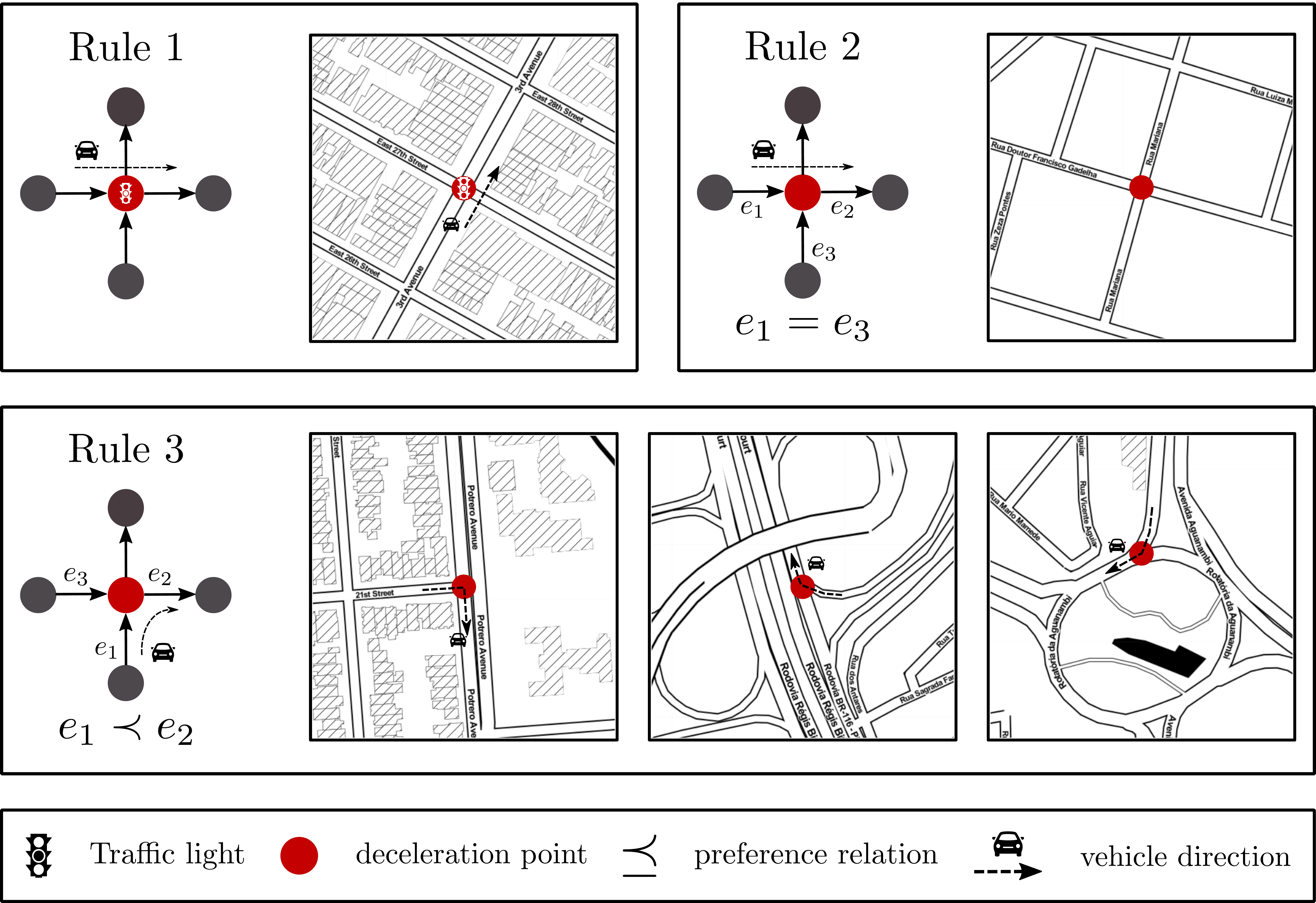}
      \caption{{\bf Rules to define Deceleration Points (DP).} Rule 1 states that any node with a traffic light is a \emph{DP}. The other rules look at preferential streets, to define which streets have priority to allow for vehicle passage and are represented by ``$\prec$'' and ``$=$'' symbols. Rule 2 establishes that if a vehicle hits a node with a segment $e_1$ with equal priority among the incident segments (equal to $e_3$), then this node gets classified as \emph{DP}, $i.e.$ $e_1 = e_3$. Rule 3 states that a node will be \emph{DP}, $i.e.$ $e_1 \prec e_2$ if the car reaches a node coming from a segment with lower preference $e_1$ than the node used to continue $e_2$.The mapping figures present examples for how rules define \emph{DP} when the vehicle approaches an avenue, roundabout or street junctions.}

\label{fig:rules}
\end{figure*}

It is possible to find $DP$ through a simulation of trips within a city. The preferential relation between intersecting streets depends on the street typology imported from \textit{OpenStreetMap}: service $\prec$ residential $\prec$ tertiary $\prec$ secondary $\prec$ primary $\prec$ trunk $\prec$ motorway. Where B $\prec$ A means that street segment A has priority over lane B and; if C $\prec$ B then C $\prec$ A. There is no preferential difference between lanes of a same typology, in that case A $=$ B. Figure \ref{fig:rules} shows the three rules applied to identify street \emph{DP}. We note that the \emph{DP} are not fixed since they depend on the direction of the vehicle when passing a specific node and can not be identified in a city without selecting a segment and direction. A certain intersection may be a \emph{DP} for one route and not for another route if the vehicle is passing through coming from a different direction, as shown in Figure \ref{fig:mapa}. Figure \ref{fig:rules} shows a simple schematic of the rules considering an existing city street topology.

\subsection{Simulating Routes}

Defining \emph{DP}s is simultaneous to the calculus of time and distance for several optimal routes simulated from graph G. In G, we simulated multiple routes with different origin points. The segments $e$ have a weight $w(e)$ equivalent to the average time it takes to travel the segment. That time is calculated as $w(e) = l(e) / V(e)$ where $l(e)$ is the segment length and $V(e)$ is the maximum street speed permitted. Since \textit{OpenStreetMap} does not inform the speed for every single segment and street, we employ a method to input missing speeds which is detailed in Supplementary Materials. A node $v$ also gets assigned a weight $w_v(v)$ to represent the time a vehicle remains stopped at an intersection when it has to stop or slow down. Nodes classified as DPs have $w_v(v) = 10$ (in seconds), otherwise $w_v(v) = 0$. We discuss how a variation in  $w_v$ impacts the results in the Supplementary Material.

The segment sizes depend on a time threshold ($\tau$) which defines the maximum amount of time a vehicle can take to read a destination, in other words, the sum of the weights of nodes and edges must be smaller or equal to the specified $\tau$. We adapt the algorithm from \cite{dijkstra1959note} to obtain the optimal routes (the pseudo-code is described in the Supplementary Material). In \cite{dijkstra1959note}, the algorithm stops running when it finds the optimal path between two pre-defined points. In our adaptation, the algorithm runs a shortest path from a random origin point during a time limit $\tau$. Thus, there is no established final point and time is the limiting factor for the extent of the route as well as the nodes and segments weights.

Figure \ref{fig:ilustracao}(a) shows a point of origin $O_1$ chosen to simulate routes for a specific city region. We simulate all optimal (fastest) routes that can be traveled starting from $O_1$ during a specified $\tau$. The green color exemplifies routes that start from an origin point and reach several destination points. The edge points show the furthest destinations reached to shape an area that represent the region which can be accessed starting from $O_1$ considering the time limit $\tau$. When connected, the edge points form an isochrone area \cite{biazzo2019general}. Such final points are not necessarily placed in city intersections and can also be points in the middle of a block since they must equate to places where an optimal route has weight equal to $\tau$. When an edge coordinate is not an intersection, we can conduct a linear interpolation in the node to find the edge coordinate.

The time needed to reach one node from a city's network ($t_d$, where $d$ is the node id) is calculated and stored from a single execution of the Dijkstra algorithm. Once we obtain the values of $t_d$ for a given origin and a value for $\tau$ we can find the edge points. We generate 20 $\tau$ values for each city ($\tau_1, \tau_2, \tau_3, ..., \tau_{20}$), where $\tau_1=1$ minute and $\tau_{20}=max(t_d)$ represent the amount of time it takes to get to a city's boundary from an initial point. The other values for $\tau$'s ($\tau_2, \tau_3, ..., \tau_{19}$) are generated to occupy a logarithmic scale. Each value for $\tau_i$ represents an area that can be covered in the city, starting from a shorter time until getting to an area that covers the entire city. The figure \ref{fig:ilustracao}(b) presents the various isochrones generated for an origin point $O_1$ and with each of the 20 values for $\tau$.

\subsection{Relation between Time and Distance}

\begin{figure}[h!]
\centering
    \includegraphics[width=0.39\textwidth]{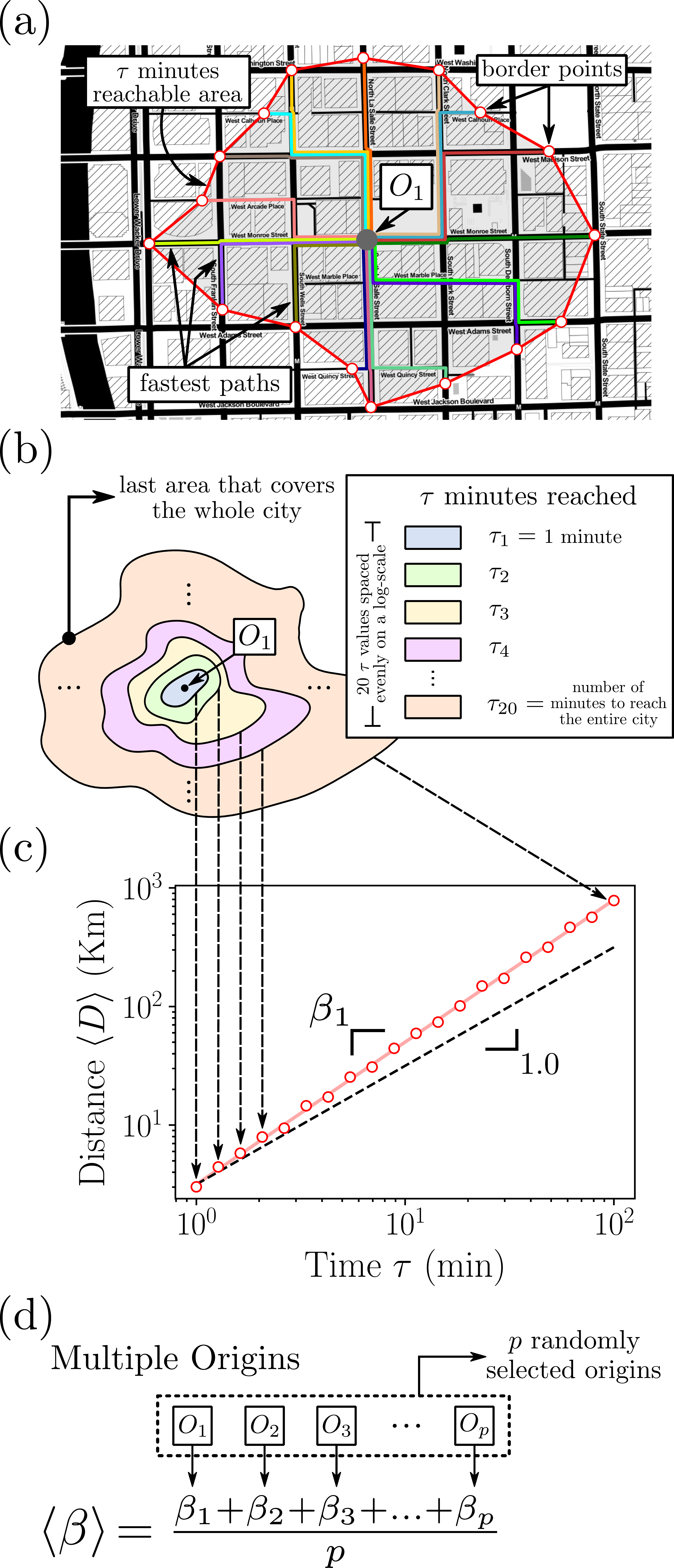}
     \caption{{\bf Method to estimate a city's $\beta$} In (a) we show a city region where we simulated multiple optimal routes from all directions from a point of origin $O_1$. The routes are represented by different colors and take a shorter time than $\tau$. From those points we define an area that can be reached in $\tau$ minutes, represented by the area outlined in red. In (b), for a same origin $O_1$ we use multiple values of $\tau$. The legend in (b) indicates how we chose $\tau$ values. (c) shows how we establish a correlation between $\tau$ in the x-axis and the average distance reached $\langle D \rangle$ for the optimal fastest routes in the y-axis, both in logarithmic scale. Each point of this x-y relation is associated to an area of (b), where the value from the y-axis is defined as the average distance of the optimal routes within an area and the value from the x-axis is the $\tau$ used to reach that area. The solid red line shows the regression with a better fit between those points, with an inclination of $\beta_1$. The dashed line in black is a guideline, with exponent equal to $1.0$. In (d) we define the calculus used to obtain a city's $\langle \beta \rangle$, which is the average of all $\beta_i$ values. Each $\beta_i$ is associated to a point of origin $O_i$ where the number of points defined is $p$ (cf, Fig. 3d).}

\label{fig:ilustracao}
\end{figure}

It is possible to establish a relation between the time and distance traveled for each origin point. Figure \ref{fig:ilustracao}(c) shows the relation between time and distance for point $O_1$. Each point in the graph illustrates an isochrone area obtained from the varying $\tau$ values. The $x$ axis represents the time for optimal routes, in minutes, which is also equal to the $\tau$ for that area, and the $y$ axis represents the average distance of the optimal routes within the area, $\langle D \rangle$, in Km, given a single unique distance value for each time limit. The axis are plotted in log-log scale. The correlation between $\tau$ and $\langle D \rangle$ helps to understand how routes with different lengths take place within a city.

The relation between time and distance is formally described by a Power Law, 
\begin{equation}
\langle D \rangle=a\tau^\beta
\end{equation}
\noindent where $\tau$ is time, $\langle D \rangle$ quantifies the distance run, $a$ is a pre-factor and $\beta$ is the exponent we want to measure.

The linear regression exponent between $log(\langle D \rangle)$ and $log(\tau)$ shows the efficiency of longer paths simulated within a city. If a correlation coefficient is larger than $1$, then a trip two times longer, for example, will reach an average distance more than twice as long, indicating a gain of scale.

In the example from figure \ref{fig:ilustracao}(c) the $\beta_1$ exponent for the correlation between time and distance is associated to origin $O_1$, showing how $\beta$ gets calculated from a single origin. The exponent value $\langle \beta \rangle$ for the entire city with multiple $p$ origin points, is the average of all $\beta_i$, where $i=1,2,3,..,p$.

\section{Results}

\begin{figure}[h!]
\centering
    \includegraphics[width=0.45\textwidth]{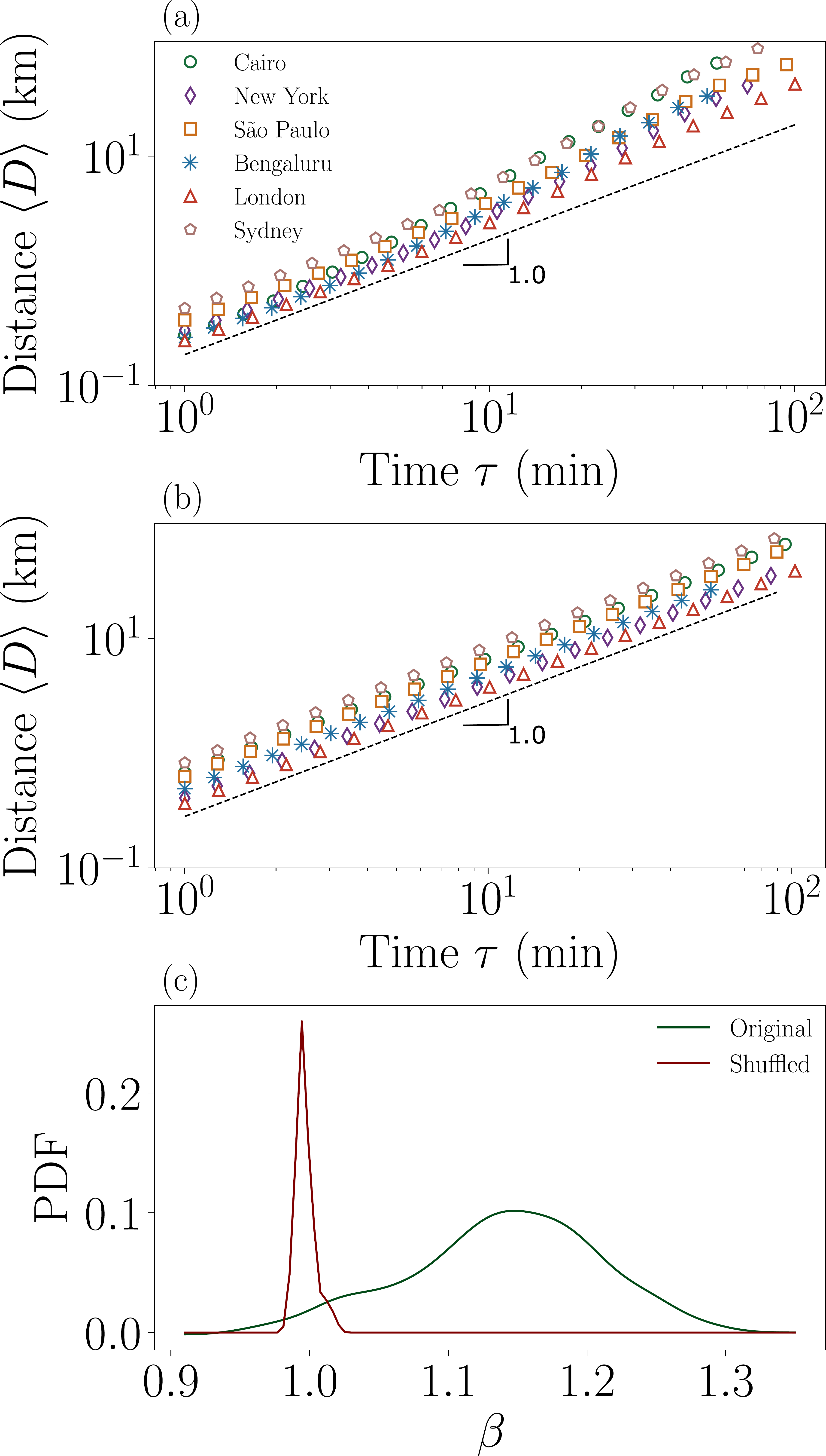}
     \caption{{\bf Correlation between time and distance and the distribution of $\langle \beta \rangle$}. In (a) we show the six average curves of the correlation between time $\tau$ in $x$, and distance $\langle D \rangle$ in $y$, representing different cities. In (b) we show the same graph as in (a) but we scrambled the location of the DP, with exponents described by $\beta_s$. The $\beta_s$ for these six cities are: Cairo: 1.40, New York: 1.24, São Paulo: 1.17, Bengaluru: 1.26, London: 1.14, and Sydney: 1.24. We added dashed lines to the (a) and (b) plots to represent an exponent of $1.0$. In (c) the full line represents the Nadaraya–Watson non-parametric regression of the $\langle \beta \rangle$ and $\langle \beta_s \rangle$ histograms.}

\label{fig:alometria5}
\end{figure}

\begin{figure*}[h!]
\centering
    \includegraphics[width=1\textwidth]{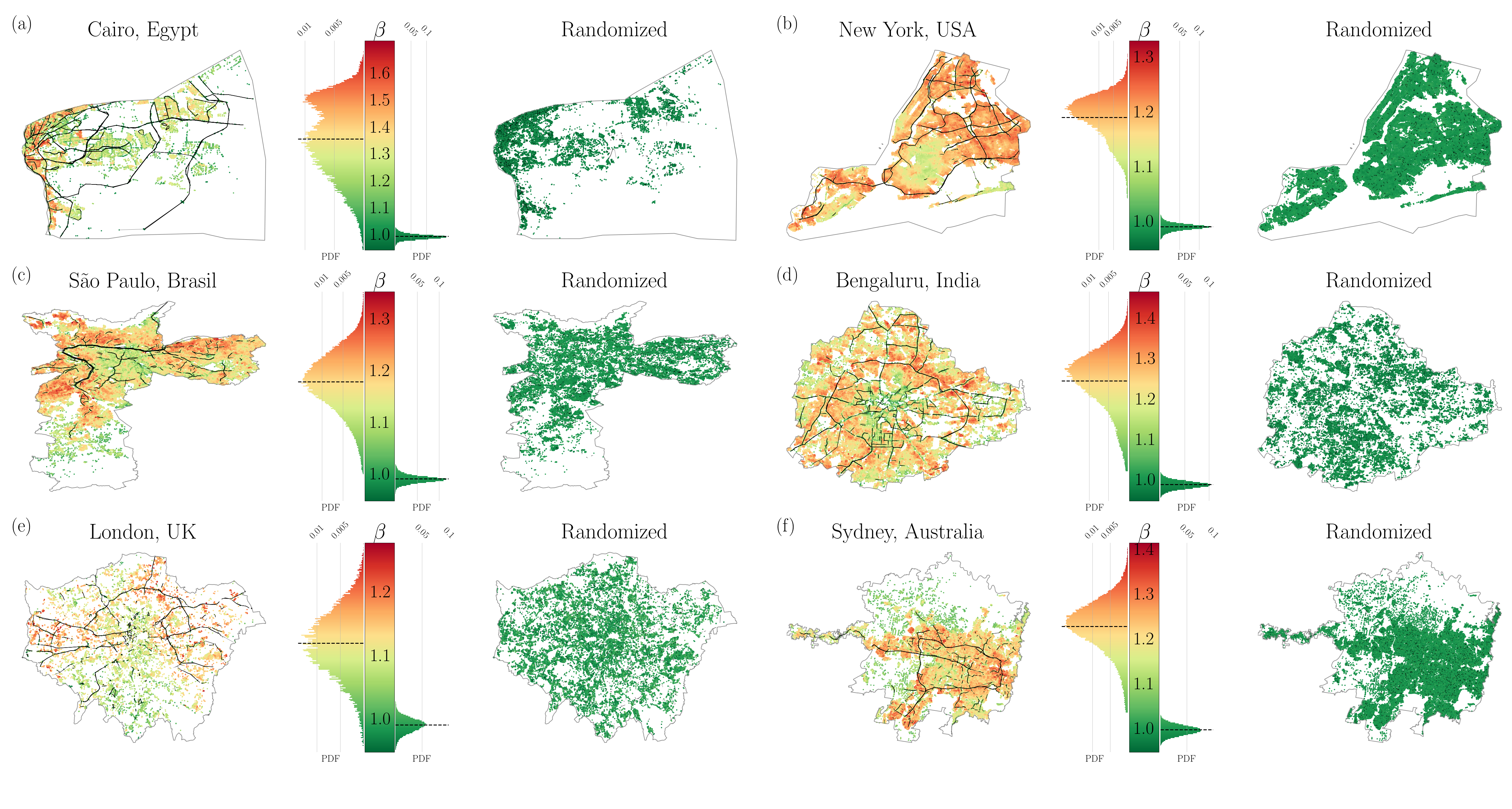}
     \caption{{\bf Spatial distribution of $\beta's$ for different cities}.In (a-e), we calculate the values for $\langle \beta \rangle$ (on the left) and $\langle \beta_s \rangle$ (on the right) for each of those six cities. We take all the nodes from the street network as origin points. The points are colored according to their exponent value and their color is painted by the color scheme at the center. The black lines represent the larger SWDP. The function for probability density for each experiment's $\langle \beta_s \rangle$ is shown in both sides of the color scale to represent each result from the execution, and a dashed line also shows its average distribution. This figure also shows the linearity of the $\langle \beta_s \rangle$ exponent for a random experiment, which stays around 1 for all cities. It also shows how SWDPs are important to ensure the non-linear characteristic of $\langle \beta_s \rangle$'s, indicating there is a spatial correlation between values with high exponents and SWDP.}

\label{fig:map5}
\end{figure*}
This section verifies the correlation between the time threshold $\tau$, and the average distances that are reachable $\langle D \rangle$ for the cities selected. As previously stated, this paper encompasses 228 cities from different continents and calculates a $\langle \beta \rangle$ value for each. Table \ref{Tab:graphs_info} details the non-linear relation between time and distance, measured by $\langle \beta \rangle$ , largely superior to $1.0$.

We selected six mega-cities from different continents for a detailed analysis. Figure \ref{fig:alometria5}(a) presents the regression of $\tau$ for $\langle D \rangle$, showing a non-linear relation for all cities when compared with the dashed line of reference ($\langle \beta \rangle=1$).

\subsection{The role of the spatial distribution of DPs and SWDPs}

To understand how the spatial structure of the DPs and SWDP impact the  $\langle \beta \rangle$ exponent, we conducted an experiment by doing a random scrambling of of the $w_v(v)$ and $w(e)$ values. As a result, we created a completely random signaling structure to re-run the simulation for the optimal routes for each city. The scrambling process was conducted for every node triplex. A triplex $A \rightarrow B \rightarrow C$ is a route between two nodes $A \rightarrow B$ e $B \rightarrow C$. The node $A \rightarrow B$ represents the current street where the vehicle is driving and $B \rightarrow C$ indicates the street where the vehicle wants to go to. Depending on the priority of the streets $A \rightarrow B$ and $B \rightarrow C$, the node $B$ can be categorized as a \emph{DP} or not according to the rules detailed in Figure \ref{fig:rules}. After classifying which triplex are \emph{DP} or not, these classifications are scrambled to erase the spatial correlation between the \emph{DP} in the city. The time $\tau$ and distances $\langle D \rangle$ were run again to calculate a new $\langle \beta_s \rangle$ exponent for the random configuration of the \emph{DP}. We conducted a similar process to randomly distribute the street speeds $i.e.$, which also altered the node weights $w(e)$.

Figure \ref{fig:alometria5}(b) shows the regression results for for the same cities when we scramble the \emph{DP} and street speeds, revealing that the non-linear aspect evident in (a) disappears. The experiment shows that the non-linear relation between $\tau$ and $\langle D \rangle$ is a consequence of the spatial correlation that exists between the street network and \emph{DP}. Figure \ref{fig:alometria5}(c) shows the kernel density estimation for the distribution of $\langle \beta \rangle$ and $\langle \beta_s \rangle$ for all 228 cities. When we compare the histograms, we see that most cities have $\langle \beta_s \rangle$ greater than 1.0 and that the non-linear relation disappears with the scrambling experiment.

Figure \ref{fig:map5} shows the spatial distribution of the values of $\beta$ for each of the six cities presented in Figures \ref{fig:alometria5} (a) and (b). To create this example, we use all the nodes for each city as origin points and execute the proposed methodology to associate a $\beta$ to each node, represented by the node color. We executed the experiment twice for each city. On the first run, the routes simulated took into account the city signaling (shown to the left) and the second run we employed the scrambling process previously explained (shown to the right). We observe that the effects of scrambling were: i) to break continuity of free flow continuous street segments and (ii) to make the distribution of  $\langle \beta \rangle$ exponents more homogeneous, close to 1 in every city. Both effects can be witnessed on the maps to the right.

\subsection{The role of the large SWDPs}

Figure \ref{fig:map5} shows the more extensive (\emph{SWDP}), highlighted in black. In this paper we selected the 20\% longest \emph{SWDP} of each city. Next to the color bars we illustrate the density probability functions of the $\beta$ to the left and $\beta_s$ to the right. The results presented on the images to the left reveal the heterogeneity of the $\beta$ values, depending on the initial point where routes are simulated. Such heterogeneity is explained by the fact that our method simulates routes from multiple random points of origin to estimate the $\langle \beta \rangle$ for each city.

We notice there are origin points which produce routes with high $\beta$ around large \emph{SWDP}. In New York, for instance, \emph{Brooklyn} concentrates most of the smaller exponent values and practically no large \emph{SWDP}, while that scenario is completely opposite for \emph{Queens} and \emph{Bronx}. Cairo is the city where such correlation can be the hardest to visualize since it is the only city where the distribution of $\beta$ values is multimodal. Still, it is possible to notice that hot spots concentrate to the left along with most of the large \emph{SWDP}. In general, Figure \ref{fig:map5} shows that the large SWDP explain the non-linearity since they boost the increase in speed and allow for a gain of scale by reducing time for longer trips within cities.

\begin{figure}[h!]
\centering
    \includegraphics[width=0.5\textwidth]{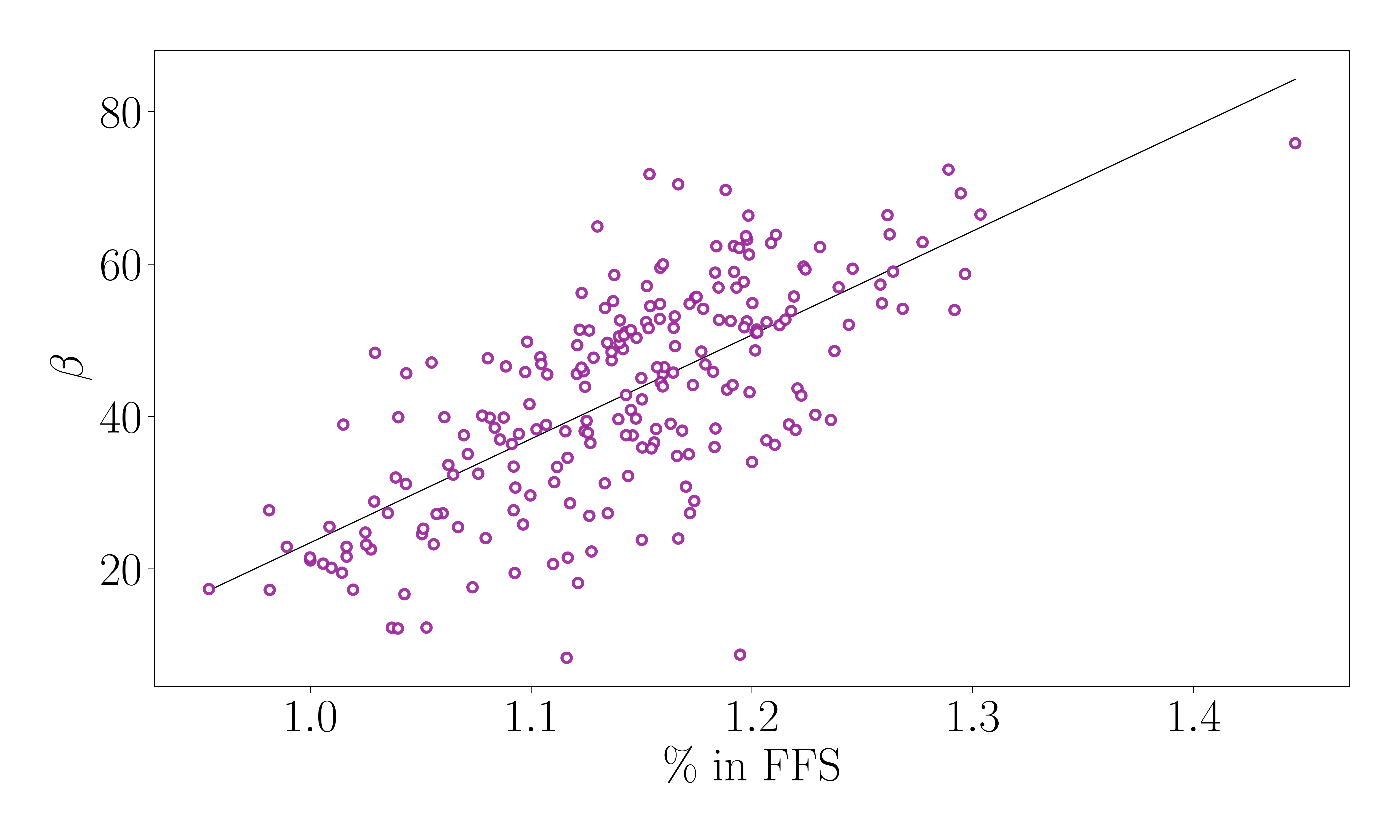}
     \caption{{\bf Correlation between $\delta^{\%}_{SWDP}$ and $\beta$}. Each point represents a city where $\delta^{\%}_{SWDP}$ is in the $x$ axis and the $\delta^{\%}_{SWDP}$ is in the $y$ axis. The black line shows a linear regression that establishes a relation between both axis.}
\label{fig:percxbeta}
\end{figure}

Figure \ref{fig:percxbeta} shows another evidence of how \emph{SWDP} are determining to the phenomena of non-linearity presented in this paper. We correlate the $\beta$ exponent with the percentage of trips completed in \emph{SWDP} for each city ($\delta^{\%}_{SWDP}$). This percentage illustrates how much gain of scale one can achieve with a route based on two factors. First, the greater the \emph{SWDP}, the greater the gain of scale for a route in a trip. Second, the frequency with which trips use \emph{SWDP} also increases the possibility of a gain of scale, and, in turn, an increase in $\beta$. Thus, we observe from figure \ref{fig:percxbeta} that the higher the density of $\delta^{\%}_{SWDP}$, the more $\beta$ is exponentially higher for a specific city.

\begin{figure}[h!]
\centering
    \includegraphics[width=0.48\textwidth]{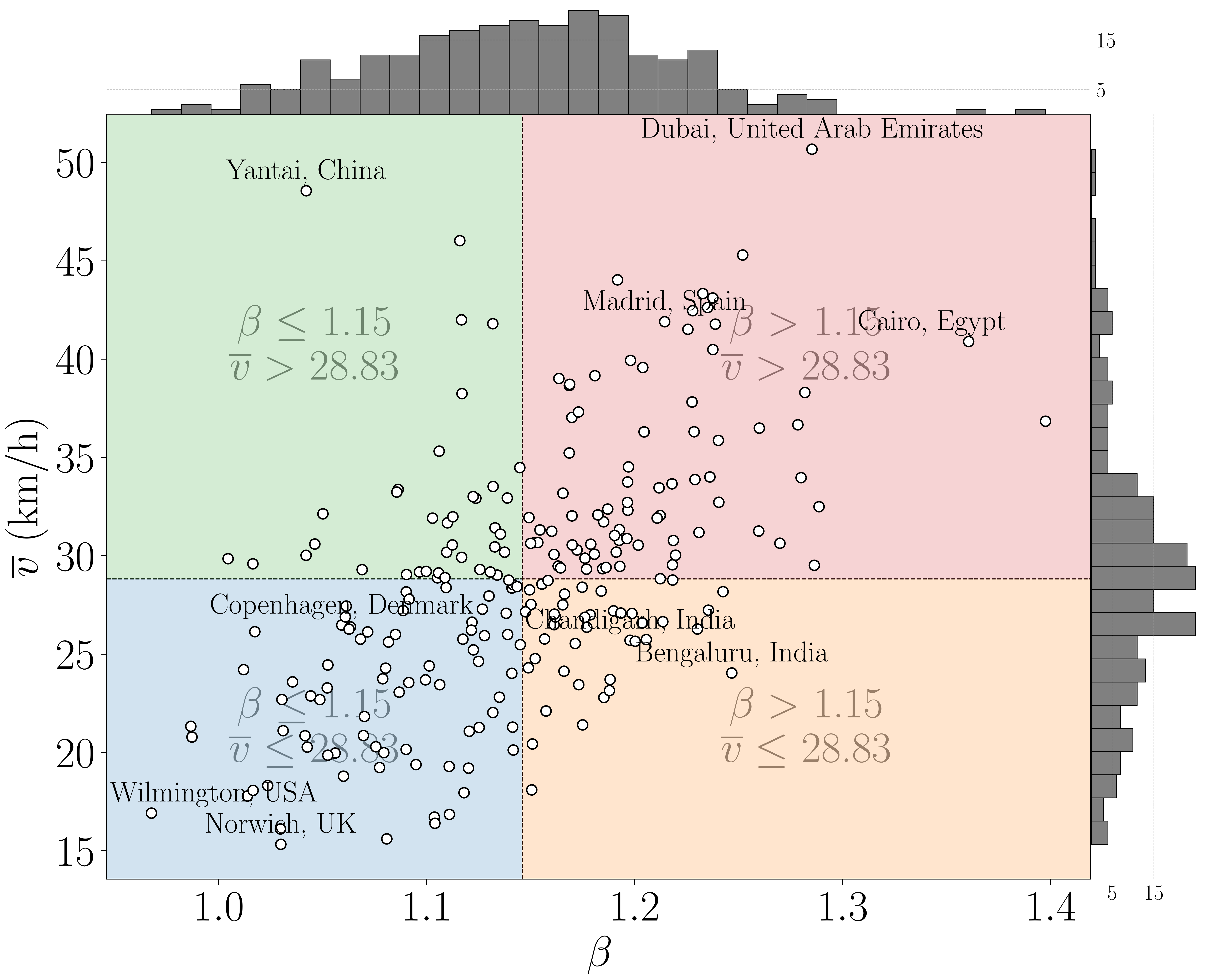}
     \caption{{\bf $\beta$ exponent versus average speed}. Each point represents a city where, the $x$ axis presents the average exponent ($\beta$) and the $y$ axis presents the average speed $\overline{v}$ achieved for each trip simulated. The histograms for $\beta$ and $\overline{v}$ values are shown in the upper and right-hand corner, respectively. The graph was divided in four different colored quadrants according to their average $\beta$ and $\overline{v}$ values.}
\label{fig:grid}
\end{figure}

\begin{table*}[p!]
\caption{Exponent for all cities simulated}
\centering
\scriptsize
\begin{tabular}{ll|ll|ll}
\hline
City & $\langle \beta \rangle$ & City &  $\langle \beta \rangle$ & City &  $\langle \beta \rangle$  \\ 
\hline
\hline

Accra, Ghana & 1.13 $ \pm 0.02$ & Adelaide, Australia & 1.23 $ \pm 0.02$ & Akureyri, Iceland & 1.05 $ \pm 0.02$ \\ 
Albuquerque, USA & 1.19 $ \pm 0.02$ & Amsterdam, Netherlands & 1.29 $ \pm 0.02$ & Antwerp, Belgium & 1.26 $ \pm 0.03$ \\ 
Aracaju, Brasil & 1.16 $ \pm 0.02$ & Asunción, Paraguay & 1.12 $ \pm 0.02$ & Atlanta, USA & 1.12 $ \pm 0.02$ \\ 
Austin, USA & 1.2 $ \pm 0.02$ & Baltimore, USA & 1.07 $ \pm 0.01$ & Bamako, Mali & 1.16 $ \pm 0.02$ \\ 
Barcelona, Spain & 1.14 $ \pm 0.02$ & Baton Rouge, USA & 1.15 $ \pm 0.02$ & Beira, Mozambique & 1.12 $ \pm 0.02$ \\ 
Belgrade, Serbia & 1.14 $ \pm 0.02$ & Bengaluru, India & 1.25 $ \pm 0.02$ & Bergen, Vestland, Norway & 1.22 $ \pm 0.02$ \\ 
Bern, Switzerland & 1.21 $ \pm 0.03$ & Birmingham, UK & 1.09 $ \pm 0.01$ & Bogotá, Colombia & 1.17 $ \pm 0.02$ \\ 
Boston, USA & 1.06 $ \pm 0.02$ & Boulder, USA & 1.09 $ \pm 0.02$ & Brisbane, Australia & 1.18 $ \pm 0.02$ \\ 
Bristol, UK & 1.08 $ \pm 0.02$ & Brussels, Belgium & 1.17 $ \pm 0.02$ & Bucharest, Romania & 1.14 $ \pm 0.02$ \\ 
Budapest, Hungary & 1.13 $ \pm 0.02$ & Buenos Aires, Argentina & 1.16 $ \pm 0.02$ & Buffalo, New York, USA & 1.09 $ \pm 0.02$ \\ 
Cairo, Egypt & 1.36 $ \pm 0.02$ & Calgary, Canadá & 1.24 $ \pm 0.02$ & Cali, Colombia & 1.13 $ \pm 0.02$ \\ 
Cambridge, UK & 1.03 $ \pm 0.01$ & Campinas, Brasil & 1.2 $ \pm 0.03$ & Cancun, Mexico & 1.11 $ \pm 0.04$ \\ 
Caracas, Venezuela & 1.18 $ \pm 0.03$ & Cartagena, Colombia & 1.19 $ \pm 0.02$ & Casablanca, Morocco & 1.28 $ \pm 0.02$ \\ 
Cayenne, France & 1.01 $ \pm 0.02$ & Chandigarh, India & 1.2 $ \pm 0.02$ & Charlotte, USA & 1.13 $ \pm 0.02$ \\ 
Chicago, USA & 1.15 $ \pm 0.02$ & Christchurch, New Zealand & 1.09 $ \pm 0.02$ & Ciudad del Este, Paraguay & 1.15 $ \pm 0.02$ \\ 
Cologne, Germany & 1.24 $ \pm 0.02$ & Columbus, USA & 1.18 $ \pm 0.02$ & Conakry, Guinea & 1.21 $ \pm 0.02$ \\ 
Copenhagen, Denmark & 1.06 $ \pm 0.01$ & Cork, Ireland & 1.22 $ \pm 0.03$ & Curitiba, Brasil & 1.16 $ \pm 0.02$ \\ 
Dallas, USA & 1.2 $ \pm 0.02$ & Dar es Salaam, Tanzania & 1.11 $ \pm 0.02$ & Denver,USA & 1.12 $ \pm 0.02$ \\ 
Detroit, USA & 1.18 $ \pm 0.03$ & Dresden, Germany & 1.09 $ \pm 0.02$ & Dubai, United Arab Emirates & 1.29 $ \pm 0.02$ \\ 
Dublin, Ireland & 1.05 $ \pm 0.01$ & Edinburgh, UK & 1.12 $ \pm 0.02$ & Edmonton, Canadá & 1.2 $ \pm 0.02$ \\ 
Eugene, Oregon, USA & 1.08 $ \pm 0.02$ & Fes, Morocco & 1.23 $ \pm 0.02$ & Florianópolis, Brasil & 1.19 $ \pm 0.03$ \\ 
Fortaleza, Brasil & 1.17 $ \pm 0.02$ & Frankfurt, Germany & 1.15 $ \pm 0.02$ & Geneva, Switzerland & 0.99 $ \pm 0.01$ \\ 
Genoa, Italy & 1.16 $ \pm 0.02$ & Georgetown, Guyana & 1.02 $ \pm 0.02$ & Goiânia, Brasil & 1.17 $ \pm 0.02$ \\ 
Graz, Austria & 1.06 $ \pm 0.02$ & Guadalajara, Mexico & 1.14 $ \pm 0.02$ & Guayaquil, Ecuador & 1.24 $ \pm 0.03$ \\ 
Hamilton, New Zealand & 1.11 $ \pm 0.02$ & Hanoi, Vietnam & 1.19 $ \pm 0.02$ & Ho Chi Minh City, Vietnam & 1.25 $ \pm 0.04$ \\ 
Hong Kong & 1.17 $ \pm 0.02$ & Honolulu, USA & 1.13 $ \pm 0.02$ & Houston, USA & 1.2 $ \pm 0.02$ \\ 
Innsbruck, Austria & 1.06 $ \pm 0.02$ & Iquitos, Peru & 0.99 $ \pm 0.02$ & Jacksonville, Florida, USA & 1.19 $ \pm 0.02$ \\ 
Jersey City, USA & 1.01 $ \pm 0.01$ & Joensuu, Finland & 1.13 $ \pm 0.02$ & Joinville, Brasil & 1.17 $ \pm 0.02$ \\ 
João Pessoa, Brasil & 1.14 $ \pm 0.02$ & Kampala, Uganda & 1.04 $ \pm 0.02$ & Kansas City,USA & 1.16 $ \pm 0.02$ \\ 
Kolkata, India & 1.08 $ \pm 0.02$ & Kuala Lumpur, Malaysia & 1.24 $ \pm 0.03$ & Kyoto, Kyoto Prefecture, Japan & 1.16 $ \pm 0.01$ \\ 
La Plata, Argentina & 1.04 $ \pm 0.01$ & Las Vegas, USA & 1.24 $ \pm 0.02$ & Leipzig, Germany & 1.09 $ \pm 0.01$ \\ 
Lisboa, Portugal & 1.22 $ \pm 0.02$ & Liverpool, UK & 1.15 $ \pm 0.02$ & Ljubljana, Slovenia & 1.22 $ \pm 0.03$ \\ 
London, UK & 1.12 $ \pm 0.02$ & Londrina, Brazil & 1.17 $ \pm 0.02$ & Los Angeles, USA & 1.18 $ \pm 0.02$ \\ 
Louisville, USA & 1.17 $ \pm 0.02$ & Luanda, Angola & 1.07 $ \pm 0.02$ & Lyon, France & 1.06 $ \pm 0.01$ \\ 
Maceió, Brasil & 1.13 $ \pm 0.02$ & Madrid, Spain & 1.21 $ \pm 0.02$ & Manchester, UK & 1.11 $ \pm 0.02$ \\ 
Manila, Philippines & 1.14 $ \pm 0.02$ & Maputo, Mozambique & 1.19 $ \pm 0.03$ & Mar del Plata, Argentina & 1.09 $ \pm 0.02$ \\ 
Marrakesh, Morocco & 1.2 $ \pm 0.02$ & Marrakesh. Morocco & 1.2 $ \pm 0.02$ & Marseille, France & 1.16 $ \pm 0.03$ \\ 
Medellín, Colombia & 1.12 $ \pm 0.02$ & Memphis, USA & 1.17 $ \pm 0.02$ & Mexico City, Mexico & 1.21 $ \pm 0.03$ \\ 
Miami, USA & 1.15 $ \pm 0.02$ & Milan, Italy & 1.06 $ \pm 0.01$ & Milwaukee, USA & 1.1 $ \pm 0.02$ \\ 
Minneapolis, USA & 1.09 $ \pm 0.02$ & Mombasa, Kenya & 1.11 $ \pm 0.02$ & Montevideo, Uruguay & 1.13 $ \pm 0.02$ \\ 
Montreal, Canadá & 1.12 $ \pm 0.02$ & Mumbai, India & 1.28 $ \pm 0.03$ & Munich, Germany & 1.15 $ \pm 0.02$ \\ 
Mérida, Yucatán, Mexico & 1.2 $ \pm 0.02$ & N'Djamena, Chad & 1.06 $ \pm 0.02$ & Nagpur, India & 1.15 $ \pm 0.02$ \\ 
Nairobi, Kenya & 1.13 $ \pm 0.02$ & Nantes, France & 1.04 $ \pm 0.02$ & Nashville, USA & 1.17 $ \pm 0.02$ \\ 
Natal, Brasil & 1.19 $ \pm 0.03$ & New Delhi, India & 1.18 $ \pm 0.02$ & New Orleans, USA & 1.16 $ \pm 0.02$ \\ 
New York, USA & 1.19 $ \pm 0.02$ & Newark, USA & 1.03 $ \pm 0.02$ & Nice, France & 1.13 $ \pm 0.02$ \\ 
Niteroi, Brasil & 1.07 $ \pm 0.03$ & Norwich, UK & 1.03 $ \pm 0.01$ & Nottingham, UK & 1.1 $ \pm 0.02$ \\ 
Nouakchott, Mauritania & 1.13 $ \pm 0.02$ & Nouméa, New Caledonia & 1.07 $ \pm 0.02$ & Nur-Sultan, Kazakhstan & 1.12 $ \pm 0.02$ \\ 
Nuremberg, Germany & 1.17 $ \pm 0.02$ & Oakland, California, USA & 1.2 $ \pm 0.03$ & Oklahoma City, USA & 1.2 $ \pm 0.02$ \\ 
Orlando, USA & 1.16 $ \pm 0.02$ & Oslo,Norway & 1.13 $ \pm 0.03$ & Ostrava, Czechia & 1.12 $ \pm 0.02$ \\ 
Oxford, UK & 1.1 $ \pm 0.02$ & Papeete, France & 1.02 $ \pm 0.02$ & Paris, France & 1.15 $ \pm 0.02$ \\ 
Philadelphia, USA & 1.14 $ \pm 0.02$ & Phoenix, USA & 1.18 $ \pm 0.02$ & Pittsburgh, Pennsylvania, USA & 1.08 $ \pm 0.02$ \\ 
Playa del Carmen, Mexico & 1.22 $ \pm 0.02$ & Pori, Finland & 1.11 $ \pm 0.02$ & Port Harcourt, Nigeria & 1.04 $ \pm 0.01$ \\ 
Portland, USA & 1.18 $ \pm 0.02$ & Porto Alegre, Brasil & 1.11 $ \pm 0.02$ & Porto Velho, Brasil & 1.19 $ \pm 0.02$ \\ 
Porto, Portugal & 1.19 $ \pm 0.03$ & Prague, Czechia & 1.2 $ \pm 0.02$ & Pyongyang, North Korea & 1.05 $ \pm 0.02$ \\ 
Quebec City, Canadá & 1.23 $ \pm 0.03$ & Rabat, Morocco & 1.15 $ \pm 0.02$ & Raleigh, USA & 1.12 $ \pm 0.02$ \\ 
Reading, USA & 1.05 $ \pm 0.01$ & Recife, Brasil & 1.16 $ \pm 0.02$ & Reykjavik, Iceland & 1.27 $ \pm 0.03$ \\ 
Ribeirão Preto, Brazil & 1.23 $ \pm 0.02$ & Richmond, USA & 1.1 $ \pm 0.02$ & Rijeka, Croatia & 1.08 $ \pm 0.02$ \\ 
Rio de Janeiro, Brasil & 1.28 $ \pm 0.03$ & Saint Louis, Illinois, USA & 1.17 $ \pm 0.02$ & Salt Lake City, USA & 1.05 $ \pm 0.03$ \\ 
Salvador, Brasil & 1.26 $ \pm 0.03$ & Salzburg, Austria & 1.03 $ \pm 0.01$ & San Antonio, USA & 1.19 $ \pm 0.02$ \\ 
San Diego, USA & 1.23 $ \pm 0.02$ & San Miguel de Allende, Mexico & 1.09 $ \pm 0.03$ & Santa Fe, USA & 1.11 $ \pm 0.02$ \\ 
Santo André, Brasil & 1.12 $ \pm 0.02$ & Santos, Brasil & 1.11 $ \pm 0.01$ & Sapporo, Japan & 1.15 $ \pm 0.02$ \\ 
Sarajevo, Bosnia and Herzegovina & 1.08 $ \pm 0.01$ & Saskatoon, Canadá & 1.17 $ \pm 0.02$ & Scranton, Pennsylvania, USA & 1.02 $ \pm 0.01$ \\ 
Seattle, USA & 1.13 $ \pm 0.02$ & Seoul, South Korea & 1.29 $ \pm 0.03$ & Setúbal, Portugal & 1.11 $ \pm 0.02$ \\ 
Shenzhen, China & 1.23 $ \pm 0.02$ & Singapore & 1.24 $ \pm 0.02$ & Sofia, Bulgaria & 1.1 $ \pm 0.02$ \\ 
Sundsvall, Sweden & 1.1 $ \pm 0.03$ & Sydney, Australia & 1.23 $ \pm 0.02$ & São Bernardo do Campo, Brasil & 1.23 $ \pm 0.02$ \\ 
São Luís, Brasil & 1.19 $ \pm 0.03$ & São Paulo, Brasil & 1.18 $ \pm 0.02$ & Taipei, Taiwan & 1.18 $ \pm 0.02$ \\ 
Teresina, Brasil & 1.14 $ \pm 0.02$ & The Hague, Netherlands & 1.19 $ \pm 0.02$ & Timisoara, Romania & 1.09 $ \pm 0.02$ \\ 
Tirana, Albania & 1.14 $ \pm 0.03$ & Toronto, Canadá & 1.24 $ \pm 0.03$ & Tripoli, Libya & 1.24 $ \pm 0.03$ \\ 
Trondheim, Trøndelag, Norway & 1.21 $ \pm 0.02$ & Tulsa, USA & 1.21 $ \pm 0.02$ & Turin, Italy & 1.1 $ \pm 0.02$ \\ 
Turku, Finland & 1.09 $ \pm 0.02$ & Ulaanbaatar, Mongolia & 1.14 $ \pm 0.02$ & Umea, Sweden & 1.14 $ \pm 0.02$ \\ 
Utrecht, Netherlands & 1.2 $ \pm 0.02$ & Valencia, Spain & 1.18 $ \pm 0.02$ & Valparaíso, Chile & 1.21 $ \pm 0.03$ \\ 
Vancouver, Canadá & 1.02 $ \pm 0.01$ & Venice, Italy & 1.15 $ \pm 0.02$ & Vienna, Austria & 1.17 $ \pm 0.02$ \\ 
Vila Velha, Brasil & 1.18 $ \pm 0.03$ & Viña del Mar, Chile & 1.14 $ \pm 0.02$ & Washington, USA & 1.0 $ \pm 0.01$ \\ 
Wilmington, USA & 0.97 $ \pm 0.01$ & Winchester, UK & 1.19 $ \pm 0.02$ & Wollongong, Australia & 1.24 $ \pm 0.03$ \\ 
Wuhan, China & 1.12 $ \pm 0.02$ & Yantai, China & 1.04 $ \pm 0.01$ & Yaoundé, Cameroon & 1.04 $ \pm 0.01$ \\ 
Zagreb, Croatia & 1.11 $ \pm 0.02$ & Zaragoza, Spain & 1.2 $ \pm 0.03$ & Zürich, Switzerland & 1.07 $ \pm 0.02$ \\

\label{Tab:graphs_info}
\end{tabular}
\end{table*}

\section{Characterization of cities based on average speed}

Figure \ref{fig:grid} represents every city analyzed (white dots) in four quadrants according to their $\overline{v}$ and $\langle \beta \rangle$. 

The first blue quadrant, on the lower-left, has cities with low average $\overline{v}$ and smaller gains of scale ($\langle \beta \rangle$ low) for simulated trips. These are generally smaller cities with few large arterial roadways and streets with lower maximum permitted speeds. Several are praised for being people-friendly and employing good urban principles: Copenhagen, poster city for Yan Gehl's "Cities for people" (\cite{gehl2013cities}), Boulder, known for implementing a very successful Urban Growth Boundary, and London are located in that group. In general, the streetscape and urban planning in these cities is made up of smaller segments that do not permit as much free flow and make that gain of scale less prominent. Other cities in that quadrant include Miami and Detroit which have fairly high street connectivity and density (\cite{america2014measuring}) and Buffalo known for Frederick Law Olmsted's planned green-way system and accessible streets (\cite{kowsky2013best}).

The upper-left green quadrant also shows cities with lower scale efficiency but tend to generate trips with higher speeds for both long and shorter trips. These cities are less extensive cities in terms of built footprint like Frankfurt, La Plata, Vancouver and Budapest. These cities have connected streets that permit higher speeds, but also large city blocks and small total areas. For instance, Vancouver occupies a modest 115 kmˆ2 in comparison to Toronto (630 kmˆ2) and Ottawa (2,778 kmˆ2). La Plata, a planned city in Argentina, is divided in a series of six by six blocks cut by diagonal streets taking up only 27 kmˆ2. Cities with such a small extension do not allow for long SWDP, which explains its lower $\beta $ values.

The two other quadrants concentrate cities with larger $\beta$, where longer trips tend to be completed proportionately faster than shorter trips. Cities on the lower-right orange quadrant have lower average speeds and include places like Paris, New York and Bogota. These are larger cities known for their robust highway systems that permeate the city, which enable an economy of scale when driving further. However, these cities are also constrained within dense urban environments. This is likely due to the fact that these cities have taken strides to limit sprawl either naturally (such as NYC, located on an island) or legally (see \cite{long2015evaluating} for urban growth boundary in Portland). As a result, smaller building blocks and a dense street network lead to more DP points and can reduce overall speeds.

Finally, cities on the red quadrant have larger $\beta$ but lower average speed. It includes large mega-cities such as Rio de Janeiro, São Paulo, Los Angeles, Mexico City, Mumbai and Seoul. This group or world metropolis have populations in the millions and an urban footprint that had to expand accordingly. It comes as no coincidence that several cities that are able to achieve such higher $\beta$ are also those most unsustainable with a high GHS emission due to on-road transportation (\cite{wei2021keeping}). Other such cities include: Houston, known for having virtually no zoning laws; Dubai, built in the last two decades on a megalomaniac model of ineffective freeway systems; and Shenzhen, a new city founded in the 1980s and with one of the worst traffic conditions in China.

It is important to note some remarks about these quadrants. First, although an effort can be made to stifle city growth, the economy of scale is still ubiquitous in every case. The more a city presents large scale arterial roads with less DP, the more the $\beta $ coefficient increases. Second, we present those examples to make the coefficients tangible, but we do not intend to present and evaluate causality as to why cities are in a specific quadrant. Lastly, the quadrants are not tied to good or bad urban planning principles. A lower $\beta $ value could pose no harm if a city is accessible with a homogeneous distribution of services and opportunities. In that case, citizens are able to live comfortably with smaller trips. On the other hand, a city with no connecting transport infrastructure through \emph{SWDP}'s can be quite limiting for the transportation of major goods. What this study does intend to show is that urban expansion inevitably generates a super-linear relationship between the urban street morphology and travel times.

\section{Conclusion}

This article finds a non-linear relation between time and distance for trips simulated for large cities worldwide. The experiments reveal that city morphology allows for longer trips to take proportionally less time than shorter trips within a city setting, which supports previous studies that showed the same relation for trips between cities and countries \cite{varga2016further}.

We identify that non-linearity takes place in virtually every city studied, whether in small cities like Pori (Finland) and large mega-cities such as São Paulo (Brazil), with over 10 million inhabitants. The $\beta$ exponent for the correlation between time and distance indicates the magnitude of the efficiency for longer trips within a city when compared to shorter trips, meaning that cities with larger exponents tend to allow for proportionally faster trips the longer they are. It is worth noting that our findings do not suggest that cities with a higher exponent are more efficient than those with smaller exponents - the accurate statement is that cities with larger exponents have more efficient longer trips than its shorter ones.

We also present a method to select different points of origin to simulate vehicle trips and explain how to define \emph{SWDP} based on \emph{DP} where vehicles tend to stop or move slower. Non-linearity depends on the spatial distribution of \emph{DP} and street speeds, which create large \emph{SWDP}s. In general, cities organize in residential and commercial zones with a high density of \emph{DP} and smaller average permitted speeds, which are connected by avenues with less \emph{DP} and higher speeds. The street layout creates large \emph{SWDP}s, which are segments that do not require vehicles to stop or slow down and permit longer trips to be completed with overall higher average speeds, and consequently less time, than shorter trips. Simulations to randomly reassign the \emph{DP}s and maximum allowed street speeds made the \emph{SWDP}s cease to exist, which annulled the spatial correlation between those variables. In short, street topology determines the magnitude of the $\beta$ exponent.

This study has specific implications for urban studies and sustainable development. Planning principles in favor of walkability, small building blocks and high density emerge as good strategies the more cities seek to become “people-centered” and green. Yet, this study shows that streetways continue to grow through a network of \emph{SWDP} which makes driving longer distances more advantageous. We simulate and quantify why there is a positive association between travel distance and time, which contests the notion that sprawl is detrimental to the transportation of human resources. We show that the economy of scale when going further is a natural ubiquitous process and invite urban planners and practitioners to consider how cities as complex systems get created over time. 

This finding begs the following question: how can we deter the expansion of the urban footprint if there is an economy of scale when going further? Since current urban guidelines strive for sustainability by preserving rural open-space, cutting back on fossil fuels and discouraging private vehicles, planners must analyze the urban network carefully to attain those goals. Innovative approaches to deter sprawl and change the transit infrastructure require thinking about how scaling efficiency impacts the distribution of goods and services and people's willingness to drive. Characteristics of the built environment play a key role here. This study provides more evidence that physical changes to the transit infrastructure are adequate to increase travel times and can consequently discourage suburban development and driving. New urban planning strategies must consider this dynamic to fight against the natural urbanization process, which makes growth and going further more advantageous.

\section*{Acknowledgements}

V.F. has been supported  by FUNCAP and CNPQ (grant \\ 307565/2020-3).
C.C., C.P., L.F., and G.L.M. has been supported by FUNCAP.
H.P.M.M. acknowledges financial support from the Portuguese Foundation for Science and Technology (FCT) under Contracts no. PTDC/FIS-MAC/28146/2017 \\ (LISBOA-01-0145-FEDER-028146), UIDB/00618/2020, \\ and UIDP/00618/2020.

\bibliographystyle{elsarticle-harv}
\bibliography{riaibib}

\begin{thebibliography}{40}
\expandafter\ifx\csname natexlab\endcsname\relax\def\natexlab#1{#1}\fi
\expandafter\ifx\csname url\endcsname\relax
  \def\url#1{\texttt{#1}}\fi
\expandafter\ifx\csname doi\endcsname\relax
  \def\doi#1{\texttt{#1}}\fi
\expandafter\ifx\csname urlprefix\endcsname\relax\def\urlprefix{URL: }\fi
\expandafter\ifx\csname doiprefix\endcsname\relax\def\doiprefix{DOI: }\fi

\bibitem[{America(2014)}]{america2014measuring}
America, S.~G., 2014. Measuring sprawl.
\newline\urlprefix\url{http://www. smartgrowthamerica. org/measuring-sprawl
  (accessed Feb. 11, 2021)}

\bibitem[{Bettencourt et~al.(2007)Bettencourt, Lobo, Helbing, K{\"u}hnert, y
  West}]{bettencourt2007growth}
Bettencourt, L.~M., Lobo, J., Helbing, D., K{\"u}hnert, C., West, G.~B., 2007.
  Growth, innovation, scaling, and the pace of life in cities. Proceedings of
  the national academy of sciences 104~(17), 7301--7306.

\bibitem[{Biazzo et~al.(2019)Biazzo, Monechi, y Loreto}]{biazzo2019general}
Biazzo, I., Monechi, B., Loreto, V., 2019. General scores for accessibility and
  inequality measures in urban areas. Royal Society open science 6~(8), 190979.

\bibitem[{Boeing(2017)}]{boeing2017osmnx}
Boeing, G., 2017. Osmnx: New methods for acquiring, constructing, analyzing,
  and visualizing complex street networks. Computers, Environment and Urban
  Systems 65, 126--139.

\bibitem[{Boltze y Tuan(2016)}]{boltze2016approaches}
Boltze, M., Tuan, V.~A., 2016. Approaches to achieve sustainability in traffic
  management. Procedia engineering 142, 205--212.

\bibitem[{Calthorpe(2000)}]{calthorpe2000new}
Calthorpe, P., 2000. New urbanism and the apologists for sprawl [to rally
  discussion]. Places 13~(2).

\bibitem[{Caminha et~al.(2017)Caminha, Furtado, Pequeno, Ponte, Melo, Oliveira,
  y Andrade~Jr}]{caminha2017human}
Caminha, C., Furtado, V., Pequeno, T.~H., Ponte, C., Melo, H.~P., Oliveira,
  E.~A., Andrade~Jr, J.~S., 2017. Human mobility in large cities as a proxy for
  crime. PloS one 12~(2), e0171609.

\bibitem[{Caminha et~al.(2018)Caminha, Furtado, Pinheiro, y
  Ponte}]{caminha2018graph}
Caminha, C., Furtado, V., Pinheiro, V., Ponte, C., 2018. Graph mining for the
  detection of overcrowding and waste of resources in public transport. Journal
  of Internet Services and Applications 9~(1), 22.

\bibitem[{Caminha et~al.(2016)Caminha, Furtado, Pinheiro, y
  Silva}]{caminha2016micro}
Caminha, C., Furtado, V., Pinheiro, V., Silva, C., 2016. Micro-interventions in
  urban transportation from pattern discovery on the flow of passengers and on
  the bus network. En: 2016 IEEE International Smart Cities Conference (ISC2).
  IEEE, pp. 1--6.

\bibitem[{Crane y Chatman(2003)}]{Crane2003}
Crane, R., Chatman, D.~G., 2003. Traffic and Sprawl: Evidence from US Commuting
  from 1985-1997. Vol.~6. University of Southern California.

\bibitem[{Dijkstra et~al.(1959)}]{dijkstra1959note}
Dijkstra, E.~W., et~al., 1959. A note on two problems in connexion with graphs.
  Numerische mathematik 1~(1), 269--271.

\bibitem[{Ersoy(2016)}]{ersoy2016landscape}
Ersoy, E., 2016. Landscape ecology practices in planning: landscape
  connectivity and urban networks. Sustainable urbanization, 291--316.

\bibitem[{Ewing et~al.(2018)Ewing, Tian, y Lyons}]{Ewingetal2018}
Ewing, R., Tian, G., Lyons, T., 2018. Does compact development increase or
  reduce traffic congestion? Cities 72, 94--101.
\newline\urlprefix\url{http://www.sciencedirect.com/science/article/pii/S0264275116304498}
\newline\doiprefix\doi{https://doi.org/10.1016/j.cities.2017.08.010}

\bibitem[{Ewing y Bartholomew(2018)}]{Ewing2018}
Ewing, R.~H., Bartholomew, K., 2018. Best practices in metropolitan
  transportation planning. Routledge.

\bibitem[{Furtado et~al.(2017)Furtado, Furtado, Caminha, Lopes, Dantas, Ponte,
  y Cavalcante}]{furtado2017data}
Furtado, V., Furtado, E., Caminha, C., Lopes, A., Dantas, V., Ponte, C.,
  Cavalcante, S., 2017. A data-driven approach to help understanding the
  preferences of public transport users. En: 2017 IEEE International Conference
  on Big Data (Big Data). IEEE, pp. 1926--1935.

\bibitem[{Gallotti y Barthelemy(2014)}]{gallotti2014anatomy}
Gallotti, R., Barthelemy, M., 2014. Anatomy and efficiency of urban multimodal
  mobility. Scientific reports 4~(1), 1--9.

\bibitem[{Gallotti et~al.(2016)Gallotti, Bazzani, Rambaldi, y
  Barthelemy}]{gallotti2016stochastic}
Gallotti, R., Bazzani, A., Rambaldi, S., Barthelemy, M., 2016. A stochastic
  model of randomly accelerated walkers for human mobility. Nature
  communications 7~(1), 1--7.

\bibitem[{Gehl(2013)}]{gehl2013cities}
Gehl, J., 2013. Cities for people. Island press.

\bibitem[{Grimm et~al.(2008)Grimm, Faeth, Golubiewski, Redman, Wu, Bai, y
  Briggs}]{grimm2008global}
Grimm, N.~B., Faeth, S.~H., Golubiewski, N.~E., Redman, C.~L., Wu, J., Bai, X.,
  Briggs, J.~M., 2008. Global change and the ecology of cities. Science
  319~(5864), 756--760.

\bibitem[{Huang et~al.(2018)Huang, Ling, Wang, Zhang, Mao, Lin, y
  Wang}]{huang2018modeling}
Huang, Z., Ling, X., Wang, P., Zhang, F., Mao, Y., Lin, T., Wang, F.-Y., 2018.
  Modeling real-time human mobility based on mobile phone and transportation
  data fusion. Transportation research part C: emerging technologies 96,
  251--269.

\bibitem[{Jayasinghe et~al.(2019)Jayasinghe, Sano, Abenayake, y
  Mahanama}]{Jayasinghe2019}
Jayasinghe, A., Sano, K., Abenayake, C.~C., Mahanama, P. K.~S., 2019. A novel
  approach to model traffic on road segments of large-scale urban road
  networks. MethodsX 6, 1147--1163.
\newline\urlprefix\url{http://www.sciencedirect.com/science/article/pii/S2215016119301128}
\newline\doiprefix\doi{https://doi.org/10.1016/j.mex.2019.04.024}

\bibitem[{Kahn(2006)}]{Kahn2006}
Kahn, M.~E., 2006. The quality of life in sprawled versus compact cities.
  prepared for the OECD ECMT Regional Round, Berkeley, California, 27--28.

\bibitem[{{Khabbaz} et~al.(2012){Khabbaz}, {Fawaz}, y {Assi}}]{Khabbaz2012}
{Khabbaz}, M.~J., {Fawaz}, W.~F., {Assi}, C.~M., 2012. A simple free-flow
  traffic model for vehicular intermittently connected networks. IEEE
  Transactions on Intelligent Transportation Systems 13~(3), 1312--1326.
\newline\doiprefix\doi{10.1109/TITS.2012.2188519}

\bibitem[{Kowsky y Olenick(2013)}]{kowsky2013best}
Kowsky, F.~R., Olenick, A., 2013. The best planned city in the world: Olmsted,
  Vaux, and the Buffalo Park system. University of Massachusetts Press.

\bibitem[{Kraemer et~al.(2020)Kraemer, Yang, Gutierrez, Wu, Klein, Pigott,
  Du~Plessis, Faria, Li, Hanage, et~al.}]{kraemer2020effect}
Kraemer, M.~U., Yang, C.-H., Gutierrez, B., Wu, C.-H., Klein, B., Pigott,
  D.~M., Du~Plessis, L., Faria, N.~R., Li, R., Hanage, W.~P., et~al., 2020. The
  effect of human mobility and control measures on the covid-19 epidemic in
  china. Science 368~(6490), 493--497.

\bibitem[{Laidley(2016)}]{Laidley2016}
Laidley, T., 2016. Measuring sprawl:a new index, recent trends, and future
  research. Urban Affairs Review 52~(1), 66--97.
\newline\urlprefix\url{https://journals.sagepub.com/doi/abs/10.1177/1078087414568812}
\newline\doiprefix\doi{10.1177/1078087414568812}

\bibitem[{Liu et~al.(2017)Liu, Liu, Yuan, Duan, Fu, Xiong, Xu, y
  Wu}]{liu2017intelligent}
Liu, Y., Liu, C., Yuan, N.~J., Duan, L., Fu, Y., Xiong, H., Xu, S., Wu, J.,
  2017. Intelligent bus routing with heterogeneous human mobility patterns.
  Knowledge and Information Systems 50~(2), 383--415.

\bibitem[{Loder et~al.(2019)Loder, Ambühl, Menendez, y Axhausen}]{Loder2019}
Loder, A., Ambühl, L., Menendez, M., Axhausen, K.~W., 2019. Understanding
  traffic capacity of urban networks. Scientific Reports 9~(1), 16283.
\newline\urlprefix\url{https://doi.org/10.1038/s41598-019-51539-5}
\newline\doiprefix\doi{10.1038/s41598-019-51539-5}

\bibitem[{Long et~al.(2015)Long, Han, Tu, y Shu}]{long2015evaluating}
Long, Y., Han, H., Tu, Y., Shu, X., 2015. Evaluating the effectiveness of urban
  growth boundaries using human mobility and activity records. Cities 46,
  76--84.

\bibitem[{Louf y Barthelemy(2014)}]{louf2014congestion}
Louf, R., Barthelemy, M., 2014. How congestion shapes cities: from mobility
  patterns to scaling. Scientific reports 4~(1), 1--9.

\bibitem[{Makse et~al.(1995)Makse, Havlin, y Stanley}]{makse1995modelling}
Makse, H.~A., Havlin, S., Stanley, H.~E., 1995. Modelling urban growth
  patterns. Nature 377~(6550), 608--612.

\bibitem[{Mastroianni et~al.(2015)Mastroianni, Monechi, Liberto, Valenti,
  Servedio, y Loreto}]{mastroianni2015local}
Mastroianni, P., Monechi, B., Liberto, C., Valenti, G., Servedio, V.~D.,
  Loreto, V., 2015. Local optimization strategies in urban vehicular mobility.
  PloS one 10~(12), e0143799.

\bibitem[{Necula(2015)}]{Necula2015}
Necula, E., 2015. Analyzing traffic patterns on street segments based on gps
  data using r. Transportation Research Procedia 10, 276--285.
\newline\urlprefix\url{http://www.sciencedirect.com/science/article/pii/S2352146515002641}
\newline\doiprefix\doi{https://doi.org/10.1016/j.trpro.2015.09.077}

\bibitem[{Ponte et~al.(2018)Ponte, Melo, Caminha, Andrade~Jr, y
  Furtado}]{ponte2018traveling}
Ponte, C., Melo, H. P.~M., Caminha, C., Andrade~Jr, J.~S., Furtado, V., 2018.
  Traveling heterogeneity in public transportation. EPJ Data Science 7~(1), 42.

\bibitem[{Rutledge et~al.(2010)Rutledge, Price, Ross, Hewitt, Webb, y
  Briggs}]{rutledge2010thought}
Rutledge, D., Price, R., Ross, C., Hewitt, A., Webb, T., Briggs, C., 2010.
  Thought for food: impacts of urbanisation trends on soil resource
  availability in new zealand. En: Proceedings of the New Zealand Grassland
  Association. pp. 241--246.

\bibitem[{Sarzynski et~al.(2006)Sarzynski, Wolman, Galster, y
  Hanson}]{Sarzynski2006}
Sarzynski, A., Wolman, H.~L., Galster, G., Hanson, R., 2006. Testing the
  conventional wisdom about land use and traffic congestion: The more we
  sprawl, the less we move? Urban Studies 43~(3), 601--626.
\newline\urlprefix\url{https://journals.sagepub.com/doi/abs/10.1080/00420980500452441}
\newline\doiprefix\doi{10.1080/00420980500452441}

\bibitem[{Silvano et~al.(2020)Silvano, Koutsopoulos, y Farah}]{Silvano2020}
Silvano, A.~P., Koutsopoulos, H.~N., Farah, H., 2020. Free flow speed
  estimation: A probabilistic, latent approach. impact of speed limit changes
  and road characteristics. Transportation Research Part A: Policy and Practice
  138, 283--298.
\newline\urlprefix\url{http://www.sciencedirect.com/science/article/pii/S0965856420306066}
\newline\doiprefix\doi{https://doi.org/10.1016/j.tra.2020.05.024}

\bibitem[{Varga et~al.(2016)Varga, Kov{\'a}cs, Toth, Papp, y
  N{\'e}da}]{varga2016further}
Varga, L., Kov{\'a}cs, A., Toth, G., Papp, I., N{\'e}da, Z., 2016. Further we
  travel the faster we go. PloS one 11~(2), e0148913.

\bibitem[{Wei et~al.(2021)Wei, Wu, y Chen}]{wei2021keeping}
Wei, T., Wu, J., Chen, S., 2021. Keeping track of greenhouse gas emission
  reduction progress and targets in 167 cities worldwide. Frontiers in
  Sustainable Cities, 64.

\bibitem[{Zlatkovic et~al.(2019)Zlatkovic, Zlatkovic, Sullivan, Bjornstad, y
  Kiavash Fayyaz~Shahandashti}]{Zlatkovic2019}
Zlatkovic, M., Zlatkovic, S., Sullivan, T., Bjornstad, J., Kiavash
  Fayyaz~Shahandashti, S., 2019. Assessment of effects of street connectivity
  on traffic performance and sustainability within communities and
  neighborhoods through traffic simulation. Sustainable Cities and Society 46,
  101409.
\newline\urlprefix\url{http://www.sciencedirect.com/science/article/pii/S2210670718316676}
\newline\doiprefix\doi{https://doi.org/10.1016/j.scs.2018.12.037}

\end{thebibliography}

\end{document}